\documentstyle[preprint,revtex]{aps}

\begin{document}
\draft
\begin{title}
Metals in high magnetic field: a new universality class of Fermi liquids
\end{title}
\author{Victor M. Yakovenko\cite{*}}
\begin{instit}
Serin Physics Laboratory, Rutgers University, P.O.Box 849, Piscataway, NJ 08855
\end{instit}
\receipt{1992}
\begin{abstract}
   Parquet equations, describing the competition between superconducting
and density-wave instabilities, are solved for a three-dimensional
isotropic metal in a high magnetic field when only the lowest Landau
level is filled.  In the case of a repulsive interaction between
electrons, a phase transition to the density-wave state is found at
finite temperature.  In the opposite case of attractive interaction, no
phase transition is found.  With decreasing temperature $T$, the
effective vertex of interaction between electrons renormalizes toward a
one-dimensional limit in a self-similar way with the characteristic
length (transverse to the magnetic field) decreasing as
$\ln^{-1/6}(\omega_c/T)$ ($\omega_c$ is a cutoff).  Correlation
functions have new forms, previously unknown for conventional
one-dimensional or three-dimensional Fermi-liquids.
\end{abstract}


\narrowtext

   The behavior of an isotropic three-dimensional (3D) metal in a high
magnetic field has attracted attention of physicists for a long time.
In this system, the energy of an electron depends only on the momentum
along the magnetic field.  Thus, the system exhibits effects
characteristic of one-dimensional (1D) metals, while intrinsically it is
3D.  As an immediate consequence of this fact, it was suggested that the
system should be unstable with respect to charge- or spin-density-wave
(DW) formation \cite{A}.  Another suggestion was that the system could
remain superconducting (SC) in an arbitrarily high magnetic field, since
the 1D dispersion law still allows for a SC instability \cite{T}.  It
was pointed out already in Ref. \cite{A} that both instabilities must be
taken into account simultaneously in the so-called parquet
approximation.  The parquet equations were written correctly in Ref.
\cite{B}, where it was found that the DW solution is indeed an
asymptotic solution of the equations.  However, the equations were not
solved numerically, thus the question of when the asymptotic solution
develops remained open.  The present author previously solved
numerically similar equations for a quasi-one-dimensional conductor in
a high magnetic field and found phase transitions to the DW state in
both cases of repulsive and attractive interaction between electrons
\cite{Y1}.  Even if initially the interaction was attractive
(favorable for superconductivity), in the course of renormalization
the sign of the interaction was effectively changed giving rise to a
non-trivial DW state exhibiting the quantum Hall effect \cite{Y2}.  In
present paper, parquet equations for an isotropic 3D metal in a very
high magnetic field, with only the lowest spin-polarized Landau level
filled, are solved numerically.  It is found that in the repulsive
case a phase transition to the DW state occurs, in agreement with the
analysis given in Ref. \cite{B}.  In the case of attraction, no phase
transition occurs and the system remains a non-trivially correlated
metal to arbitrarily low temperatures.  With decreasing temperature
$T$, the system becomes progressively more one-dimensional, with the
characteristic interaction length perpendicular to the magnetic field
decreasing as $\zeta^{-1/6}$, where $\zeta=\ln(\omega_c/T)$, and
$\omega_c$ is a cutoff of the order of the cyclotron frequency.
However, the system never becomes strictly 1D.  By this reason, the SC
susceptibility grows as $\zeta^{1/3}\exp({\rm const}\zeta^{2/3})$.
This dependence is neither characteristic of 1D, nor of conventional
3D metals.  Thus, the system is a new non-trivial example of an
unconventional Fermi-liquid, neither of the Landau, nor of the
Luttinger type \cite{H}.  It should be emphasized also that there
exists a model where the competition between SC and DW channels is
suppressed and, thus superconductivity indeed may survive in an
arbitrarily high magnetic field \cite{L}.  However, this is not the
case for the model considered below.

   Let us consider an isotropic 3D metal with an effective mass $m$
and a parabolic dispersion law.  We assume that magnetic field is so
high that only the lowest Landau level is filled and all electrons are
spin-polarized.  Dispersion law has the form $\varepsilon=p_z^2/2m$,
where $p_z$ is the momentum along the field.  The Fermi surface
consists of the two Fermi points $p_z=\pm p_F=\pm2\pi^2\hbar nl_H^2$,
where $n$ is the volume concentration of electrons and $l_H=(\hbar
c/eH)^{1/2}$ is the magnetic length.  Coordinates $X,\;Y$ in the plane
perpendicular to the field will be measured in units of $l_H$.  The
wave functions of the lowest Landau level are \cite{LL}: \FL
\begin{equation}
\psi_x(X,Y)=\pi^{-1/4}l_H^{-1/2}\exp(ixY-(x-X)^2/2).
\label{psiX}
\end{equation}
Another representation can be obtained making a superposition of states
(\ref{psiX}):
\begin{eqnarray}
&&\psi_y(X,Y)=(2\pi)^{-1/2}\int dx\,\exp(-ixy)\psi_x(X,Y)\nonumber\\
&&=\pi^{-1/4}l_H^{-1/2}\exp(iX(Y-y)-(y-Y)^2/2).
\label{psiY}
\end{eqnarray}
To be distinguished from the running variables $X$ and $Y$, the
quantum numbers $x$ and $y$, which label the wave functions (\ref{psiX})
and (\ref{psiY}), are denoted everywhere by the lowercase letters.

   Let us introduce now the operators $\hat{a}^+(x,p_z)$ and
$\hat{b}^+(y,p_z)$ which create electrons in the eigenstates
(\ref{psiX}) and (\ref{psiY}) with $p_z$ close to $+p_F$ and $-p_F$,
respectively.  Near the Fermi energy, the electron spectrum can be
linearized in $p_z$ and the Hamiltonian of the model can be written in
the form $\hat{H}=\hat{H}_0+\hat{H}_1$, where \widetext \FL
\begin{eqnarray}
&&\hat{H}_0=2\pi l_H\int\frac{dp_z}{2\pi\hbar}v_F[(p_z-p_F)
\int dx\,\hat{a}^+(x,p_z)\hat{a}(x,p_z)
\nonumber\\
&&-(p_z+p_F)\int dy\,\hat{b}^+(y,p_z)\hat{b}(y,p_z)],\;\;\;\;\;\;
v_F=p_F/m,\label{H0} \\
&&\hat{H}_1=\int g\frac{dp_1\,dp_2\,dp_3}{(2\pi\hbar)^3}
dx\,dy\,dx'dy'\exp(-ixy'+iyx')\gamma_0(x-x',y-y')\nonumber\\
&&\hat{a}^+(x,p_1)\hat{b}^+(y,p_2)\hat{b}(y',p_3)\hat{a}(x',p_1+p_2-p_3),
\label{H1} \\
&&\gamma_0({\bf r-r'})=\exp(i{\bf r}\wedge{\bf r'})\int dX\,dY\,
\psi^*_x(X,Y)\psi^*_y(X,Y)\psi_{y'}(X,Y)\psi_{x'}(X,Y)\nonumber\\
&&=\exp(-({\bf r}-{\bf r'})^2/2).\label{g0}
\end{eqnarray} \narrowtext
In Eq. (\ref{g0}), ${\bf r}=(x,y),\;{\bf r'}=(x',y')$ and ${\bf
r}\wedge{\bf r'}=xy'-yx'$.  The function $\gamma_0$ is a form-factor of
the interaction between electrons in the representation given in
(\ref{psiX}) and (\ref{psiY}).  The interaction amplitude is
$g=g_2-g_1$, where $g_2$ and $g_1$ are the amplitudes of forward and
backward scattering.  We assume that the original
interaction is not completely local, so that $g_1\neq g_2$ and $g\neq0$;
however, its spatial range is much shorter than $l_H$.

The so-called parquet diagrams, which consist of the electron-hole and
electron-electron loops inserted into each other in all possible ways,
are the most important many-body corrections to the interaction vertex
$\gamma_0({\bf r})$ \cite{A}, \cite{B}.  The corrections
form a series in powers of \FL
$$
\xi=(|g|/(2\pi)^3v_Fl_H^2)\ln[\omega_c/{\rm max}(T,\omega,v_F||p_z|-p_F|)],
$$
where $\omega$ is a frequency, since both of the one-loop diagrams
are logarithmically divergent.  The
renormalized vertex of interaction $\gamma({\bf r},\xi)$ obeys the
following equation, shown graphically in Fig.1: \FL
\begin{eqnarray}
&&\partial\gamma({\bf r},\xi)/\partial\xi=\int d^2{\bf r}'
\gamma({\bf r}',\xi)\gamma({\bf r}-{\bf r}',\xi)
(1-e^{i{\bf r}\wedge{\bf r}'}),\label{g}\\
&&\gamma({\bf r},0)={\rm sign}(g)\gamma_0({\bf r}).\label{sign}
\end{eqnarray}
Eq. (\ref{g}) can be obtained from Eq. (2) of Ref. \cite{B} via a
Fourier transformation over the variable $k_x$.

The r.h.s. of Eq. (\ref{g}) is a difference of two terms.  The first
term is the contribution of the electron-hole loop, the second term
--- of the electron-electron loop.  If only one of these terms is
retained in the r.h.s., that corresponds to a ladder approximation
instead of a parquet one, then Eq. (\ref{g}) can be solved
analytically.  If one neglects the second term, then the equations are
diagonalized via two-dimensional (2D) Fourier transformation
$\gamma({\bf r},\xi)\rightarrow\Gamma({\bf k},\xi)$ \cite{B}:
\begin{equation}
\Gamma({\bf k},\xi)=({\rm sign}(g)\Gamma_0^{-1}({\bf k})-\xi)^{-1},
\label{gk}
\end{equation}
where $\Gamma_0({\bf k})$ is the 2D Fourier transform of
$\gamma_0({\bf r})$ (\ref{g0}).  In the case $g>0$ (repulsion), solution
(\ref{gk}) is called a ``moving pole'' \cite{GD}, because the position
$\xi_p({\bf k})=\Gamma_0^{-1}({\bf k})$ of the pole singularity in $\xi$
depends on the value of ${\bf k}$.  It follows from (\ref{g0}) that the
minimum value of $\xi_p({\bf k})$, which is equal to $\xi_c=1/2\pi$, is
attained at ${\bf k}_c=0$.  As was shown by a calculation of the
appropriate susceptibility \cite{B}, the moving pole singularity
(\ref{gk}) indicates a phase transition to a DW state, where the
densities of the electron charge and spin are modulated along the
magnetic field with a wave vector $2p_F/\hbar$, and are homogeneous in
the perpendicular plane (because ${\bf k}_c=0$).  Once the moving pole
(\ref{gk}) develops, the second term in Eq. (\ref{g}) indeed can be
neglected, because this term contains an integration over ${\bf k}$
which makes it less singular than the first term \cite{B}.

   Solving numerically the full equation (\ref{g}) at $g>0$, we
find the moving pole singularity occuring at ${\bf k}_c=0$ and
$\xi_c=1.3/2\pi$.  The value of $\xi_c$ is related to the transition
temperature by the formula
$T_c=\omega_c\exp(-(2\pi)^3\xi_cv_Fl_H^2/|g|)$.  Thus, the only effect
of the second term in Eq. (\ref{g}) in the repulsive case is a certain
decrease of the transition temperature.  Otherwise, the ladder
approximation gives qualitatively correct results.

   If the first term in the r.h.s. of Eq. (\ref{g}) is neglected, then
the resulting equation can be solved by performing the Fourier
transformation $\gamma(x_1,y,\xi)\rightarrow\lambda(x_1,x_2,\xi)$ over
the variable $y$ and introducing the function
$h(x_1,x_2,\xi)=\lambda(x_1-x_2,x_1+x_2,\xi)$ obeying the
following equation \cite{Y1}: \FL
\begin{equation}
\partial h(x_1,x_2,\xi)/\partial\xi=-\int dx\,h(x_1,x,\xi)h(x,x_2,\xi).
\label{dh}
\end{equation}
With initial conditions (\ref{sign}), Eq. (\ref{dh}) has a solution: \FL
\begin{equation}
h(x_1,x_2,\xi)=\sqrt{2\pi}\exp(-x_1^2-x_2^2)/({\rm sign}(g)+\pi\xi).
\label{h}
\end{equation}
In the case $g<0$ (attraction), Eq. (\ref{h}) has a pole
singularity at $\xi_c=1/\pi$ that indicates a phase transition to a
SC state.  Solution (\ref{h}) is called a ``standing pole'' \cite{GD},
because the position of the pole in $\xi$ does not depend on any
continuous variable.  For this reason, when expression (\ref{h}) is
substituted into the full Eq. (\ref{g}), the first term has the same
singularity as the second, thus the SC ladder approximation can
never be justified \cite{GD}.  This fact explains why it is
important to solve numerically the full Eq. (\ref{g}) in the case $g<0$.

   Since the initial vertex $\gamma_0({\bf r})$ (\ref{g0}) depends only
on $r$, which is the absolute value of ${\bf r}$, then the same holds
for $\gamma({\bf r},\xi)$.  Eq. (\ref{g}) can be rewritten for a new
function $\bar{\gamma}(r,\xi)=\gamma({\bf r},\xi)$, which depends on one
spatial argument: \FL
\begin{eqnarray}
&&\partial\bar{\gamma}(r,\xi)/\partial\xi=8\int^\infty_0r_1dr_1
\int_{|r-r_1|}^{r+r_1}r_2dr_2\,\bar{\gamma}(r_1,\xi)\bar{\gamma}(r_2,\xi)
\nonumber\\
&&\frac{\sin^2([4(r_1r_2)^2-(r_1^2+r_2^2-r^2)^2]^{1/2}/4)}
{[4(r_1r_2)^2-(r_1^2+r_2^2-r^2)^2]^{1/2}}, \label{bar}\\
&&\bar{\gamma}(r,0)={\rm sign}(g)\exp(-r^2/2).\label{bar0}
\end{eqnarray}
The numerical solution of Eq. (\ref{bar})--(\ref{bar0}) for the case
$g<0$ is shown in Fig. 2 for several values of ``time'' $\xi$.  After a
short initial evolution, the function $\bar{\gamma}(r,\xi)$ attains the
form $\bar{\gamma}_c(w(\xi)r)$, where $w(\xi)$ is a monotonically
growing function of $\xi$.  The ansatz
$\bar{\gamma}(r,\xi)=\bar{\gamma}_c(w(\xi)r)$ is consistent with Eq.
(\ref{bar}) provided $w$ is sufficiently large.  In this case, the sine
in Eq. (\ref{bar}) can be replaced by its argument, and Eq. (\ref{bar})
decouples into two equations: \FL
\begin{eqnarray}
&&dw(\xi)/d\xi=Aw^{-5}(\xi), \label{w}\\
&&2A\rho\partial\gamma_c(\rho)/\partial\rho=\int^\infty_0\rho_1d\rho_1
\int_{|\rho-\rho_1|}^{\rho+\rho_1}\rho_2d\rho_2\,
\bar{\gamma}_c(\rho_1)\bar{\gamma}_c(\rho_2) \nonumber \\
&&[4(\rho_1\rho_2)^2-(\rho_1^2+\rho_2^2-\rho^2)^2]^{1/2},\label{dgc}
\end{eqnarray}
$A$ is a constant.  It follows from (\ref{w}) that
$w(\xi)=[A(\xi-\xi_0)]^{1/6}$. This dependence indeed was found
numerically with $\xi_0=0$.  The function
$\bar{\gamma}_c(\rho)$ is also known numerically: with the convention
$A=1$, $\bar{\gamma}_c(\rho)=\bar{\gamma}(\rho\xi^{-1/6},\xi)$, where
for the latter function one can take any of the plots in Fig. 2,
except one corresponding to $\xi=0$.  In summary, the solution of the
parquet equations for the case $g<0$ has the self-similar form
\begin{equation}
\gamma({\bf r},\xi)=\gamma_c({\bf r}\xi^{1/6}),\label{gc}
\end{equation}
where we introduced a function of two variables $\gamma_c({\bf
r})=\bar{\gamma}_c(r)$.  Eq. (\ref{gc}) is neither a moving, nor a
standing pole; it is rather a ``squeezing'' solution: the effective
range of interaction varies as $\ln^{-1/6}(\omega_c/T)$, thereby making
the system increasingly one-dimensional as temperature is reduced.

   In order to calculate susceptibilities, let us add to the Hamiltonian
(\ref{H0})--(\ref{H1}) the fictitious external fields, $f_{SC}$ and
$f_{DW}$, which create electron-electron and electron-hole pairs: \FL
\begin{eqnarray}
&&\hat{H}_2=\int\frac{dp_z}{2\pi}dx\,dy\,[f_{SC}(x,y)
\hat{a}^+(x,p_z)\hat{b}^+(y,-p_z)+\nonumber\\
&&f_{DW}(x,y)\exp(-ixy)\hat{a}^+(x,p_z)\hat{b}(y,p_z-2p_F)]
+{\rm h.c.}\label{H2}
\end{eqnarray}
Let us start with the SC susceptibility.  According to the parquet
rules \cite{B}, it is necessary to calculate first a vertex
$\Psi(x,y,\xi)$, which is determined by the graphical equation shown in
Fig. 3a: \FL
\begin{eqnarray}
&&\partial\Psi({\bf r},\xi)/\partial\xi=-\int d{\bf
r}'\, \gamma({\bf r}-{\bf r}',\xi)\Psi({\bf r}',\xi)
e^{-i{\bf r}\wedge{\bf r}'},\label{dPsi}\\
&&\Psi({\bf r},0)=f_{SC}({\bf r}).
\label{Psi0}
\end{eqnarray}
To solve this equation let us make a Fourier transformation over the
variable $y$: $\Psi(x_1,y,\xi)\rightarrow\Lambda(x_1,x_2,\xi)$ and
$f_{SC}(x_1,y)\rightarrow F_{SC}(x_1,x_2)$; then, introduce the new
variables: $\Xi(\bar{x},x,\xi)=\Lambda(\bar{x}-x,\bar{x}+x,\xi)$ and $
\Phi_{SC}(\bar{x},x)= F_{SC}(\bar{x}-x,\bar{x}+x)$.
In the new variables, Eq. (\ref{dPsi}) reads:
\begin{eqnarray}
&&\partial\Xi(\bar{x},x,\xi)/\partial\xi=-\int dx'\,
\Xi(\bar{x},x',\xi)h(x',x,\xi),\label{dXi}\\
&&\Xi(\bar{x},x,0)=\Phi_{SC}(x,\bar{x}), \label{Xi0}
\end{eqnarray}
where the function $h$ was introduced earlier. Eq. (\ref{dXi}) is
diagonal and degenerate with respect to
$\bar{x}$, the center-of-mass coordinate of a Cooper pair,
so this variable can be omitted. Taking into account Eq.
(\ref{gc}), one can rewrite Eq. (\ref{dXi}) in the form: \FL
$$
\frac{\partial\Xi(x,\xi)}{\partial\xi}=-\int\frac{dx'}{\xi^{1/3}}
\lambda_c(x',\frac{2x}{\xi^{1/6}}+\frac{x'}{\xi^{1/3}})
\Xi(x+\frac{x'}{\xi^{1/6}},\xi),
$$
where $\lambda_c(x_1,x_2)$ is the Fourier transform of
$\gamma_c(x_1,y)$ over the variable $y$. An approximate solution is: \FL
\begin{eqnarray}
&&\Xi(x,\xi)=\Phi_{SC}(x)\exp(\frac{3\Gamma_0\xi^{2/3}}{2}-
\frac{2\Gamma_1x^2\xi^{1/3}}{3}), \label{Xi}\\
&&\Gamma_0=-\int^\infty_02\pi r\bar{\gamma}_c(r)dr=3.1,\nonumber\\
&&\Gamma_1=-\int^\infty_0\pi r^3\bar{\gamma}_c(r)dr=1.1.\nonumber
\end{eqnarray}
The contribution to the free energy $\delta_1F$ due to the external
field $f_{SC}$ is shown graphically in Fig. 3b:
\begin{equation}
\delta_1F=2\int^\xi_0d\zeta\int dx|\Xi(x,\zeta)|^2/|g|.\label{D1F}
\end{equation}
Substituting Eq. (\ref{Xi}) into Eq. (\ref{D1F}), one finds the
susceptibility with respect to creation of an electron pair with the
relative distance between electrons equal to $x$: \FL
\begin{equation}
\chi(x,\xi)=\xi^{1/3}\exp(3\Gamma_0\xi^{2/3}-
\Gamma_1x^2\xi^{1/3}/3)/|g|\Gamma_0.\label{xSC}
\end{equation}
When $\xi\rightarrow\infty$, the susceptibility diverges following an unusual
law $\xi^{1/3}\exp({\rm const}\xi^{2/3})$.  This behavior can be
understood qualitatively in the following way. The total amplitude of
scattering to all possible channels,
$$
\gamma_t(\xi)=\int d^2{\bf r}\,\gamma({\bf r},\xi)=-\Gamma_0/\xi^{1/3},
$$
goes to zero with increasing $\xi$. However, the function
$-2\int^\xi_0\gamma_t(\zeta)d\zeta$, which appears in the exponent of
(\ref{xSC}), increases as $\xi^{2/3}$. At the same
time, the characteristic distance between electrons in a Cooper pair
in (\ref{xSC}) squeezes as $\xi^{-1/6}$, which indicates the
increasingly one-dimensional character of the system.

Following an analogous procedure, one can find a contribution to the
free energy due to $f_{DW}$: \FL
$$
\delta_2F=\int^\xi_0\frac{d\zeta}{|g|}d^2{\bf k}|F_{DW}({\bf k})|^2
\exp\left(2\int^\zeta_0\Gamma_c(\frac{{\bf k}}{\eta^{1/6}})
\frac{d\eta}{\eta^{1/3}}\right),
$$
where $F_{DW}({\bf k})$ and $\Gamma_c({\bf k})$ are the 2D Fourier
transforms of $f_{DW}({\bf r})$ and $\gamma_c({\bf r})$.  For all values
of ${\bf k}$ the DW susceptibilities have finite limits at zero
temperature.

   The lowest order diagram, which gives a logarithmic correction to
the one-particle Green function, is shown in Fig. 4.  Renormalized
Green functions $G_{\pm}$, where the indices $\pm$ refer to the
$\hat{a}$ and $\hat{b}$ fermions respectively, can be found in
the framework of a scaling hypothesis \cite{MS}: \FL
\begin{eqnarray}
&&G_{\pm}(p_z,\omega)=\frac{\exp(-3\Gamma_2\xi^{2/3}|g|/4(2\pi)^2v_Fl_H^2)}
{\omega\mp v_F(p_z\mp p_F)},\label{Gpm}\\
&&\Gamma_2=\int^\infty_02\pi r\bar{\gamma}_c^2(r)dr=1.9.\nonumber
\end{eqnarray}
Eq. (\ref{Gpm}) differs from expressions known for
the 3D Landau and the 1D Luttinger Fermi-liquids \cite{H}.

   In conclusion, the self-similar ``squeezing'' solution (\ref{gc})
of the parquet equations has never appeared previously in condensed
matter physics.  The correlation functions (\ref{xSC}) and (\ref{Gpm})
represent a new universality class of Fermi-liquid behavior.  All of
this enriches our intuition beyond the Luttinger and the Landau
schemes in consideration of what happens when many 1D channels
interact \cite{BYA}.

The author is grateful to E. Abrahams for the support of this work via
the NSF Grant No. DMR 89-06958.

\figure{ Parquet equations for the vertex of interaction
$\gamma(x_2-x_1,y_2-y_1,\xi)$ shown as a filled circle. The bare vertex
$\gamma_0(x_2-x_1,y_2-y_1)$ is denoted as a dot. The
Green functions of electrons with momenta $p_z$ close to $\pm p_F$
are shown as solid and dotted lines. \label{Fig.1}}

\figure{ The vertex of interaction $\bar{\gamma}(r,\xi)$ as
a function of $r$ at $2\pi\xi=0$ (curve a), 5 (b), 500 (c) and 3200
(d). \label{Fig.2}}

\figure{ a) Parquet equations for the vertex $\Psi(x,y,\xi)$
shown as a filled triangle. The external field $f_{SC}(x,y)$
is denoted as a square. b) Contribution to the free energy $\delta_1F$ due
to the field $f_{SC}$. \label{Fig.3}}

\figure{ The lowest order logarithmic correction to a Green function.
\label{Fig.4}}

\end{document}